\documentclass[12pt,oneside]{amsart}

\usepackage{amsmath, amssymb, amsthm, amscd, url}

\newcommand{\la}{{\lambda}}

\newcommand{\er}{\eqref}

\newcommand{\ad}{{\rm ad\,}}

\newcommand{\msl}{\mathfrak{sl}}
\newcommand{\mg}{\mathfrak{g}}

\newcommand{\mh}{\mathfrak{h}}

\newcommand{\we}{\mathfrak{w}\!\mathfrak{e}}

\newcommand{\ev}{\mathrm{ev}}

\newcommand{\Com}{\mathbb{C}}

\newcommand{\pd}{\partial}

\newtheorem{theorem}{Theorem}

\newtheorem{lemma}{Lemma}
\newtheorem{corollary}{Corollary}

\theoremstyle{definition}
\newtheorem{definition}{Definition}
\newtheorem{example}{Example}
\newtheorem{remark}{Remark}

\begin{document}

\author{Sergey A. Igonin}
\address{Department of Mathematics \\ 
Utrecht University \\
P.O. Box 80010 \\ 
3508 TA Utrecht \\ 
the Netherlands} 

\email{igonin@mccme.ru}

\title[Miura transformations and homogeneous spaces]
{Miura type transformations and homogeneous spaces}

\date{}



\begin{abstract}
We relate Miura type transformations (MTs) over an evolution system 
to its zero-curvature representations with values in Lie algebras $\mg$. 
We prove that certain homogeneous spaces of $\mg$ produce MTs 
and show how to distinguish these spaces. 
For a scalar translation-invariant evolution equation this allows to classify all MTs 
in terms of homogeneous spaces of the Wahlquist-Estabrook algebra of the equation. 
For other evolution systems this allows to construct some MTs. 
As an example, we study MTs over the KdV equation, a 5th order equation of Harry-Dym type, 
and the coupled KdV-mKdV system of Kersten and Krasilshchik.  
\end{abstract}

\maketitle

\section{Introduction}

In this paper we study $(1+1)$-dimensional evolution systems 
\begin{gather}
\label{sv}
\frac{\pd v^i}{\pd t}=R^i(x,t,v^1,\dots,v^k,v^1_1,\dots,v^k_1,\dots,v^1_r,\dots,v^k_r),\\
\label{su}
\frac{\pd u^i}{\pd t}=P^i(x,t,u^1,\dots,u^k,u^1_1,\dots,u^k_1,\dots,u^1_p,\dots,u^k_p),\\
\notag
v^i_j=\frac{\pd^j v^i}{\pd x^j},\quad u^i_j=\frac{\pd^j u^i}{\pd x^j},\quad i=1,\dots,k,
\end{gather}
and transformations 
\begin{equation}
\label{mt}
u^i=S^i(x,t,v^1,\dots,v^k,v^1_1,\dots,v^k_1,\dots,v^1_n,\dots,v^k_n)
\end{equation}
such that if $v^1,\dots,v^k$ satisfy~\er{sv} then~\er{mt} satisfy~\er{su}.
By analogy with the classical Miura transformation connecting the KdV and mKdV equations, 
such transformations are called \emph{Miura type transformations} (MTs in short). 

It is well known that MTs play an important role in the theory of integrable evolution systems. 
Chains of MTs generate B\"acklund transformations~\cite{backl,backl1}. 

If $P^i,\,R^i,\,S^i$ do not depend on $x$ and $t$, the MT is called \emph{translation-invariant}.

In the scalar case $k=1$, the problem to find all pairs~\er{su},~\er{mt} for a given system~\er{sv} 
was solved efficiently in~\cite{diff_sub}. 
 
We study the opposite problem: for a given system~\er{su}, 
how to find all pairs~\er{sv},~\er{mt}? 
It seems that this problem was systematically studied only 
for translation-invariant MTs of the linear equation $u_t=u_3$~\cite{khab} 
and the KdV equation~\cite{drin,guil,khab}. 
In particular, a set of MTs over the KdV equation was constructed using homogeneous 
spaces of certain Lie groups (in~\cite{drin}) and loop groups (in~\cite{guil}). 
These MTs do not exhaust all translation-invariant MTs over the KdV equation, but,  
knowing all integrable equations of the form $u_t=u_3+f(u,u_1,u_2)$,  
one can show that all other MTs can be obtained from these by introduction of a potential. 

We relate MTs~\er{mt} with zero-curvature representations of~\er{su} (ZCRs in short) 
dependent on $x,\,t$, $u^i_j$, $j\le p-1$. 
It turns out that such a ZCR with values in a Lie algebra $\mg$ 
and a certain representation of $\mg$ by vector fields on a manifold $W$ determine a MT.
Among other requirements, the image of $\mg$ under the representation must span 
the tangent spaces of $W$, that is, the manifold $W$ is a homogeneous space of $\mg$.

This construction is not surprising if one recalls the theory of coverings of PDEs~\cite{nonl,cfg1}.
It is becoming clear~\cite{cfg1} that each covering is determined by a $\mg$-valued ZCR and 
a vector field representation of some Lie algebra $\mg$. 
Since MTs are a particular type of coverings, 
it remains to determine which ZCRs and representations lead to MTs. 
However, in order to be self-contained, we do not introduce the coverings terminology 
and work in local coordinates. 

If $k=1$ and \er{su} is translation-invariant, 
we prove that every MT arises in this way from the ``universal'' 
ZCR with values in the Wahl\-quist-Estabrook algebra of~\er{su}. 
This allows to reduce classification of MTs 
to the classification of certain homogeneous spaces of the Wahl\-quist-Estabrook algebra. 
As an example, we obtain that any (not necessarily translation-invariant) 
MT over the KdV equation is of order not greater than 3. 
Also, we recover the Lie groups of~\cite{drin} as the Lie groups 
associated with some quotients of the Wahlquist-Estabrook algebra 
and explain why the method of~\cite{drin} does not give all translation-invariant MTs 
over the KdV equation.

Another considered example is the equation $u_t=u^{\frac52}u_5$~\cite{hdbt,konop}. 
Using its Wahlquist-Estabrook algebra computed in~\cite{hdbt}, 
we show that any MT over this equation is of order not greater than 5 and construct 
a MT of order 3. The corresponding modified equation~\er{hd5m} 
may be a new integrable equation. 

For nonscalar systems we obtain only 
a sufficient condition for a ZCR to define a MT. 
As an example, we construct a MT over 
the coupled KdV-mKdV system of Kersten and Krasilshchik~\cite{kdv-mkdv}. 
This MT arises from the ZCR obtained in~\cite{kdv-mkdv-sak}.  
Again, the corresponding modified system may be new.

There are also more general transformations of evolution equations, 
where one changes not only dependent variables, 
but also the $x$ variable (see, e.g.,~\cite{ibr,diff_sub,sak}). 
It remains an interesting open question whether 
our Lie algebraic methods can be generalized 
for studying of these transformations. 
 
The paper is organized as follows. In Section~\ref{lie} 
we study Lie algebras actions of special type that will 
later turn out to be responsible for MTs. 
In Sections~\ref{scal},~\ref{wemt} we describe MTs of scalar 
evolution equations and their relations with Wahlquist-Estabrook algebras. 
Finally, in Section~\ref{mtsys} we study MTs of 
nonscalar evolution systems. 
 
\section{Actions of Lie algebras on manifolds}
\label{lie}

Recall that an \emph{action} of a Lie algebra $\mg$ on a manifold $W$ is
a homomorphism $\rho\colon\mg\to D(W)$ to the Lie algebra $D(W)$ of vector fields on $W$.
The action is said to be \emph{transitive} if for each point $a\in W$ the mapping 
$$
\ev_{\rho,a}\colon\mg\to T_a W,\quad g\mapsto \rho(g)_a,
$$ 
is surjective. 
%In this case $W$ is called a \emph{homogeneous space} of $\mg$, since . 
Two actions $\rho_i\colon\mg\to D(W_i),\,i=1,2$, are said to be \emph{isomorphic} 
if there is a diffeomorphism $\varphi\colon W_1\to W_2$ 
such that $\rho_2=\varphi_*\rho_1$.

Below our considerations are always local.
The results are valid in both categories of smooth and complex-analytic manifolds.
Depending on the category considered, all functions are supposed to be smooth
or complex-analytic.

In what follows we often consider a (possibly infinite) 
chain of subalgebras 
\begin{equation}
\label{ggg}
\mg^1\subset\mg^2\subset\mg^3\subset\dots\subset\mg 
\end{equation}
and an action $\rho\colon\mg\to D(W)$. 
It easy to see that there is a non-empty open subset $W_c\subset W$ 
such that 
\begin{gather}
\label{aa}
\forall\,a,\,a'\in W_c,\quad
\forall\,i\quad  
\dim\ev_{\rho,a}(\mg^i)=\dim\ev_{\rho,a'}(\mg^i),\\
\notag 
\dim\ev_{\rho,a}(\mg)=\dim\ev_{\rho,a'}(\mg).
\end{gather}
Moreover, if $W$ is connected and $\rho$ is analytic, 
one can choose $W_c$ to be dense in $W$. 
Since we study locally nondegenerate points only, 
below in the paper we always assume $W=W_c$.
 
Denote $m_i=\dim\ev_{\rho,a}(\mg^i)$ for $a\in W_c$.
Due to inclusions~\er{ggg} we have 
\begin{equation}
\label{mlem}
m_1\le m_2\le m_3\le\dots.
\end{equation}
\begin{lemma}
\label{ooo}
In the above described situation, 
suppose that there is $V\in D(W)$ such that 
for each $i$ the Lie algebra generated 
by the subspace $\rho(\mg^i)+[\rho(\mg^i),V]$ coincides with 
$\rho(\mg^{i+1})$. Set $n=\dim W$. 
Suppose that 
\begin{equation}
\label{msn}
m_1\ge s,\quad m_{n-s}<n,\quad \exists\,k\quad m_k=n. 
\end{equation}
Then 
\begin{equation}
\label{msn1}
m_i=s+i-1,\quad i=1,\dots,n-s+1.
\end{equation}
Moreover, for each point $z\in W$ on a neighborhood of $z$ 
there is a function $w$ such that 
\begin{equation*}
%\label{sysdzw}
d_zw\neq 0,\quad \rho(\mg^{n-s})(w)=0,
\end{equation*}
and this function is unique up to a change $w\mapsto g(w)$. 
Set
\begin{equation*}
%\label{sysww}
\tilde w^i=V^{i-1}(w),\quad i=1,\dots,n-s+1.
\end{equation*}
One can find functions $\tilde w^{n-s+2},\dots,\tilde w^{n}$ 
such that $\tilde w^1,\dots,\tilde w^n$ form a system of coordinates 
on a neighborhood of $z$. 
\end{lemma}
\begin{proof}
For a set $S$ of vector fields denote by $\langle S\rangle$  
the submodule (over the algebra of functions) 
of vector fields generated by $S$.  
By condition~\er{aa} and the Frobenius theorem, 
there are coordinates $w^1,\dots,w^n$ on a neighborhood of $z$ 
such that 
\begin{equation}
\label{mod}
\forall\,i\quad
\langle\rho(\mg^i)\rangle=\left\langle\frac{\pd}{\pd w^1},\dots,
\frac{\pd}{\pd w^{m_i}}\right\rangle.
\end{equation}
Denote module~\er{mod} by $M_i$.
Suppose that $m_k=m_{k+1}$ for some $k$. 
Then $[M_k,V]\subset M_k$
and, therefore, $m_p=m_k$ for all $p\ge k$. 
Combining this property with~\er{msn} and~\er{mlem}, 
we obtain~\er{msn1}. 
 
Now we can take $w=w^n$. Using the equalities
$$
M_i+\langle[M_i,V]\rangle=M_{i+1},\quad 
i=1,\dots,n-s,
$$
by induction on $i$ one proves 
$$
M_{n-s-i+1}(\tilde w^i)=0,\quad 
\frac{\pd \tilde w^i}{\pd w^{n-i+1}}\neq 0,\quad 
i=1,\dots,n-s+1,
$$ 
where 
$$
M_0=
\left\langle\frac{\pd}{\pd w^1},\dots,
\frac{\pd}{\pd w^{s-1}}\right\rangle.
$$
Therefore, if we set $\tilde w^{n-s+1+i}=w^i$ for $i=1,\dots,s-1$, 
the functions $\tilde w^1,\dots,\tilde w^n$ will 
be local coordinates on a neighborhood of $z$.
\end{proof}

\section{MTs from ZCRs of scalar evolution equations}
\label{scal}

Consider two scalar evolution equations
\begin{gather}
  \label{ee}
  u_t=P(x,t,u,u_1,\dots,u_p),\quad u_k=\frac{\pd^k u}{\pd x^k},\\
  \label{evol1}
  v_t=R(x,t,v,v_1,\dots,v_r),\quad v_k=\frac{\pd^k v}{\pd x^k},
\end{gather}
connected by a MT
\begin{equation}
  \label{sub}
  u=S(x,t,v,v_1,\dots,v_n).
\end{equation}
The maximal integer $n$ such that
\eqref{sub} depends nontrivially on $v_n$ is called the \emph{order}
of the MT. A MT obtained from this one by a substitution $v\mapsto g(v)$ is said 
to be \emph{equivalent} to the initial MT.

%We will associate with such a transformation a transitive action of some Lie algebra
%on a $n$-dimensional manifold. Some special properties of these actions allow
%to construct and classify MTs for a fixed equation \er{ee}.

Introduce new variables
\begin{equation}
\label{wv}
w^i=\frac{\pd^{i-1}v}{\pd x^{i-1}},\quad i=1,\dots,n,
\end{equation}
and rewrite system \er{evol1}, \er{sub} as follows
\begin{equation}
  \label{dsc}
  \begin{aligned}%[bb]
  \frac{\pd w^i}{\pd x}&=w^{i+1},\quad i=1,\dots,n-1,\\
  \frac{\pd w^n}{\pd x}& =a(w^1,\dots,w^n,x,t,u),\\
  \frac{\pd w^i}{\pd t}&=b^i(w^1,\dots,w^n,x,t,u,\dots,u_{p-1}),\quad i=1,\dots,n,
  \end{aligned}
\end{equation}
where $p$ is the order of \eqref{ee}.
And vice versa, it is easily seen that any consistent system of this form with 
the non-degeneracy condition
\begin{equation}
\label{du}
\exists\,w^1_0,\dots,w^n_0,x_0,t_0,u_0\quad
    \frac{\pd a}{\pd u}(w^1_0,\dots,w^n_0,x_0,t_0,u_0)\neq 0 
\end{equation}
determines a MT of order $n$ for \er{ee} as follows:
\begin{itemize}
    \item substitute \er{wv} to \er{dsc},
    \item taking into account~\er{du}, 
    from equation \er{dsc} express locally $u=S(x,t,v,v_1,\dots,v_n)$,
    \item let $D=\sum_{i\ge 0}v_{i+1}{\pd}/{\pd v_i}$, 
    then equation \er{evol1} is given by
    $$
    v_t=b^1(v,v_1,\dots,v_{n-1},x,t,S,D(S),\dots,D^{p-1}(S)).
    $$
\end{itemize}

Consider the \emph{total derivative} operators
\begin{align*}
%\label{dx}
  D_x&=\frac{\pd}{\pd x}+\sum_{j\ge 0} u_{j+1}\frac{\pd}{\pd u_j},\\
%\label{dt}
  D_t&=\frac{\pd}{\pd t}+\sum_{j\ge 0} D_x^j\bigl(P(x,t,u,u_1,\dots,u_p)\bigl)\frac{\pd}{\pd u_j}
\end{align*}
and more general overdetermined systems
\begin{equation}
  \label{dscg}
  \begin{aligned}%[bb]
  \frac{\pd w^i}{\pd x}& =a^i(w^1,\dots,w^n,x,t,u),\quad i=1,\dots,n,\\
  \frac{\pd w^i}{\pd t}&=b^i(w^1,\dots,w^n,x,t,u,\dots,u_{p-1}),\quad i=1,\dots,n,
  \end{aligned}
\end{equation}
consistent modulo \er{ee}.
Clearly, an invertible change of variables
\begin{equation}
\label{ww}
w^i\mapsto f^i(w^1,\dots,w^n)
\end{equation}
leads to a new system of the form \er{dscg}. Two systems related by such
a change of variables are said to be \emph{equivalent}.

System \er{dscg} is completely determined by the vector fields
\begin{gather*}
    A=\sum_{i=1}^n a^i(w^1,\dots,w^n,x,t,u)\frac{\pd}{\pd w^i},\\
    B=\sum_{i=1}^n b^i(w^1,\dots,w^n,x,t,u,\dots,u_{p-1})\frac{\pd}{\pd w^i}.
\end{gather*}
Consistency of \er{dscg} modulo \er{ee} is equivalent to the equation
\begin{equation}
    \label{ab}
    [D_x+A,\,D_t+B]=0.
\end{equation}

Recall that two functions
\begin{equation}
\label{zcr}
M(x,t,u),\quad N(x,t,u,\dots,u_{p-1})
\end{equation}
with values in a Lie algebra $\mg$ constitute a \emph{zero-curvature representation}
(ZCR in short) for \er{ee} if
\begin{equation}
\label{zcr1}
[D_x+M,\,D_t+N]=D_x N-D_t M+[M,N]=0.
\end{equation}
Then each action $\rho\colon\mg\to D(W)$ and a choice of local coordinates
$w^1,\dots,w^n$ in $W$ determine a consistent
system of the form \er{dscg} with $A=\rho(M)$ and $B=\rho(N)$,
since equation~\er{ab} follows from \er{zcr1}.
Clearly, different choices of coordinates in $W$ or isomorphic
actions determine equivalent systems \er{dscg}.

\begin{definition}
\label{gk}
Suppose that a ZCR~\er{zcr} is given.
For each $k\in\mathbb{N}$ we define a subalgebra $\mg^k$ of $\mg$
by induction on $k$ as follows:
\begin{itemize}
    \item $\mg^0=0$,
    \item $\mg^1$ is the subalgebra generated by all elements 
$$
M(x,t,u)-M(x',t',u')\in\mg,
$$
where $x,\,t,\,u,\,x',\,t',\,u'$ run through all admissible (real or complex) 
values of the corresponding variables. 
\item $\mg^{k+1}$ is generated by the subspaces $\mg^k$ and $[\mg^k,M(x,t,u)]$.
\end{itemize}
\begin{remark}
Note that due to the definition of $\mg^1$ the space  
$$
\mg^k+[\mg^k,M(x,t,u)]
$$ 
does not depend on the values of $x,\,t,\,u$.
\end{remark}
Set also $\tilde\mg=\cup_{k\ge 0}\mg^k$.
\end{definition}

\begin{theorem}
\label{mts} 
Suppose that system~\er{dscg} arises from an action $\rho\colon\mg\to D(W)$.
Then the following two statements are equivalent. 
\begin{enumerate}
\item There are $z=(w^1_0,\dots,w^n_0)\in W$ and 
an invertible transformation~\er{ww}  
on a neighborhood of $z$ such that system~\er{dscg} 
takes the form~\er{dsc} with~\er{du}. 
\item There are $z\in W$ and a neighborhood $W_0$ of $z$ 
such that  
\begin{gather}
\label{zmn}
\dim\ev_{\rho,z}(\tilde\mg)=n, \\
\label{amn}
\forall\,a\in W_0\quad \dim\ev_{\rho,a}(\mg^{n-1})<n,\\
\label{duth}
\exists\,x_0,\,t_0,\,u_0\quad
\frac{\pd}{\pd u}(\rho(M))(z,x_0,t_0,u_0)\neq 0.
\end{gather}
\end{enumerate}
In this case on a neighborhood of $z$ 
there is a function $w$ such that 
\begin{equation}
\label{dzw}
d_zw\neq 0,\quad \rho(\mg^{n-1})(w)=0,
\end{equation}
and it is unique up to a change $w\mapsto g(w)$. The functions
\begin{equation}
\label{w}
\tilde w^i=\rho\bigl(M(x,t,u)\bigl)^{i-1}(w),\quad i=1,\dots,n,
\end{equation}
do not depend on $x,\,t,\,u$ and are local coordinates in which 
system~\er{dscg} takes the desired form~\er{dsc},~\er{du}.
\end{theorem}

\begin{proof}
If system~\er{dscg} is of the form~\er{dsc},~\er{du} 
then by the definition of $\mg^k$ we obtain that on a neighborhood of $z$
the image of $\mg^k$ in each tangent space of $W$ 
is spanned by $\pd/\pd w^{n-k+1},\dots,\pd/\pd w^{n}$.
This obviously implies~\er{zmn} and~\er{amn}, and~\er{duth} 
follows from~\er{du}.

Conversely, let~\er{zmn},~\er{amn}, and~\er{duth} hold. 
Then existence of $w$ and the fact that $\tilde w^1,\dots,\tilde w^n$ 
are local coordinates follow from Lemma~\ref{ooo} for $s=1$ 
and $V=\rho\bigl(M(x,t,u)\bigl)$. 
In particular, the functions $\tilde w^1,\dots,\tilde w^{n-1}$ 
are invariant under $\rho(\mg^1)$.  
Combining this fact with the formula 
$\tilde w^{k+1}=\rho\bigl(M(x,t,u)\bigl)(\tilde w^k)$, 
by induction on $k$ one gets that each function $\tilde w^k$
does not depend on $x,\,t,\,u$. 
 
It is easily seen that system~\er{dscg} is of the form~\er{dsc} 
in the coordinates $\tilde w^1,\dots,\tilde w^n$.
Finally condition~\er{du} follows from~\er{duth}. 
\end{proof}

\begin{remark}
\label{trans}
In the above theorem, to construct a MT it is sufficient to know  
the restriction of $\rho$ to some neighborhood of $z$, and 
condition~\er{zmn} implies that the action $\rho|_{\tilde\mg}$ 
is transitive on a neighborhood of $z$. 
\end{remark}
\begin{corollary}
\label{ub}
If $\mg^m=\mg^{m+1}$ for some $m\ge 0$ 
\textup{(}equivalently, $\tilde\mg=\mg^m$\textup{)} then
actions of $\mg$ cannot produce MTs of order greater than $m$.
\end{corollary}
\begin{proof}
By Theorem~\ref{mts} and the above remark, a MT of order $n$
is determined by a transitive action $\rho$ of $\tilde\mg$ such that
$\rho(\mg^{n-1})$ is not transitive. Since in our case
$\tilde\mg=\mg^k$ for any $k\ge m$, there are no such actions for $n>m$.
\end{proof}

\begin{remark} 
\label{nongen}
If $M$ does not depend on $x,\,t$ then~\er{duth} follows from~\er{zmn} and \er{amn}. 
Indeed, from the above proof in this case $\dim\ev_{\rho,z}(\mg^1)=1$, 
which implies that
$$
\exists\,u_0\quad\frac{\pd\rho(M)}{\pd u}(z,u_0)\neq 0.
$$
\end{remark}
  
\section{Wahlquist-Estabrook algebras and MTs}

\label{wemt}

\subsection{General results}

For a scalar translation-invariant equation 
\begin{equation}
\label{tie}
u_t=P(u,u_1,\dots,u_p)
\end{equation}
recall the definition of the Wahlquist-Estabrook algebra~\cite{Prol} 
from the point of view of~\cite{nonl}. 
Consider the equation
\begin{equation}
\label{tizcr}
[D_x+A(u),\,D_t+B(u,u_1,\dots,u_{p-1})]=0,
\end{equation}
where $A,\,B$ are functions with values in a (not specified in advance) Lie algebra $\mg$. 
The algebra $\mg$ can also be the algebra of vector fields on a manifold $W$, 
then $A$ and $B$ are vector fields on $W$ dependent on $u,\dots,u_{p-1}$. 
Suppose that for any $\mg$ equation~\er{tizcr} implies
\begin{gather}
\label{wea}
A=\sum_{i=1}^{k_1}f_i(u)F_i,\\
\label{web}
B=\sum_{i=1}^{k_2}g_i(u,u_1,\dots,u_{p-1})G_i,
\end{gather}
where $f_i,\,g_i$ are some fixed scalar functions, which do not depend on $\mg$, 
and $F_i,\,G_i$ are elements of $\mg$. 
Moreover, suppose that functions~\er{wea} satisfy~\er{tizcr} if and only if 
some Lie algebra relations hold between the elements $F_1,\dots,F_{k_1}$, 
$G_1,\dots,G_{k_2}$. 
In this case the quotient of the free Lie algebra generated by the letters $F_i,\,G_j$ 
over these relations is called the Wahlquist-Estabrook algebra of~\er{tie} 
and is denoted~$\we$. 
Functions~\er{wea},~\er{web} constitute a ZCR with values in $\we$ such that 
any consistent translation-invariant system~\er{dscg} arises from this ZCR 
and some action of $\we$.

The Wahlquist-Estabrook algebra exists for practically all known equations 
(see, e.g.,~\cite{dodd,nonl} and references therein). 

By Theorem~\ref{mts} and Remarks~\ref{trans},~\ref{nongen},
we obtain the following result. 
\begin{theorem}
\label{mtswe}
Translation-invariant MTs of order $n$ of~\er{tie} are in one-to-one correspondence 
with actions $\rho$ of $\we$ such that $\rho(\tilde\we)$ is transitive 
and $\rho(\we^{n-1})$ is not transitive. Locally isomorphic actions 
determine equivalent MTs. 
\end{theorem}

It turns out that non-translation-invariant MTs can also be described 
in terms of actions of $\we$. By the construction of~\cite{kirn}, 
with any system~\er{dsc} we can associate the following translation-invariant system
\begin{equation*}
  \begin{aligned}%[bb]
\frac{\pd\hat w^1}{\pd x}&=\frac{\pd\hat w^2}{\pd t}=1,\quad 
\frac{\pd\hat w^2}{\pd x}=\frac{\pd\hat w^1}{\pd t}=0,\\
  \frac{\pd w^i}{\pd x}&=w^{i+1},\quad i=1,\dots,n-1,\\
  \frac{\pd w^n}{\pd x}& =a(w^1,\dots,w^n,\hat w^1,\hat w^2,u),\\
  \frac{\pd w^i}{\pd t}&=b^i(w^1,\dots,w^n,\hat w^1,\hat w^2,u,\dots,u_{p-1}),\quad i=1,\dots,n.
  \end{aligned}
\end{equation*}
This system is consistent provided that the initial system~\er{dsc} is consistent 
and equation~\er{ee} (equation~\er{tie}) is translation-invariant. 
It is determined by an action $\rho$ of $\we$ on the manifold 
with coordinates $\hat w^1,\hat w^2,w^1,\dots,w^n$. 
If system~\er{dsc} arises from a MT~\er{sub},~\er{evol1} then we have~\er{du}, 
which similarly to the proof of Theorem~\ref{mts} implies 
$$
\dim\ev_{\rho,a}(\tilde\we)=n,\quad 
\dim\ev_{\rho,a}(\we^{n-1})=n-1.
$$ 
This observation implies the following (compare with Corollary~\ref{ub}). 

\begin{theorem}
\label{xtorder}
If for some $n$ we have $\we^n=\tilde\we$ then any \textup{(}not necessarily 
translation-invariant\textup{)} MT over~\er{tie} is of order not greater than $n$. 
\end{theorem}

\subsection{MTs of the KdV equation}

Consider the KdV equation 
$$
u_t=u_3+u_1u
$$ 
in the complex-analytic category. According to \cite{kdv1}, we have 
$$
\we=H\oplus\msl_2(\mathbb{C})\otimes_\Com\Com[\lambda],
$$ 
where $H$ is the $5$-dimensional nilpotent Heisenberg algebra with
the basis $r_i,\,i=-2,-1,0,1,2$, and the commutator table
$$
[r_{-1},r_1]=[r_2,r_{-2}]=r_0,\quad [r_i,r_j]=0\ \forall\,i+j\neq 0
$$
and the Lie bracket in $\msl_2\otimes\Com[\la]$ is defined as follows 
$$
[g_1\otimes f_1(\la),\,g_2\otimes f_2(\la)]=[g_1,g_2]\otimes f_1(\la)f_2(\la),
\quad g_i\in\msl_2,\ f_i(\la)\in\Com[\la].
$$  
Below an element $g\otimes f(\la)$ of $\msl_2\otimes\Com[\la]$
will be written simply as $gf(\la)$.

The universal ZCR reads
\begin{gather}
\label{kdvM}
A(u)=X_1+\frac13uX_2+\frac16u^2X_3,\\
\label{kdvX}
X_1=r_1-\frac12y+\frac12z\la,\quad X_2=r_{-1}+z,\quad X_3=r_{-2}, 
\end{gather}
where $h,\,y,\,z$ is a basis of $\msl_2$ with the relations
$[h,y]=2y,\,[h,z]=-2z,\,[y,z]=h$.
Here the form of $B(u,u_1,u_2)$ in~\er{web} is not important for us.

From~\er{kdvM} and~\er{kdvX}, using Definition~\ref{gk}, one obtains 
\begin{gather*}
\we^1=\langle X_2,\,X_3\rangle,\quad
\we^2=\langle r_{-2},\,r_{-1},\,z,\,2r_0+h\rangle,\\
\we^3=\we^k=\tilde\we=\langle   \msl_2\otimes\Com[\la],\,r_{-2},\, r_{-1},\, r_0\rangle
\quad\forall\,k\ge 3.	
\end{gather*}
By Theorem~\ref{xtorder}, any MT of the KdV equation is of order not greater than 3.
For translation-invariant MTs this was proved in~\cite{khab}.

Let us explain how our method of constructing MTs includes the one of~\cite{drin}.
Set $\mg=\msl_2(\mathbb{C})\otimes_{\mathbb{C}}\Com[\lambda]$. 
We have the natural projection $\we\to\mg$ that maps $H$ to zero.
Combining it with the above ZCR, we obtain a ZCR 
with values in $\mg$ whose $x$-part reads 
\begin{equation*}
M(u)=-\frac12y+\frac12z\la+\frac13uz. 	
\end{equation*} 

For this ZCR we have
$$
\mg^1=\langle z\rangle,\quad\mg^2=\langle z,h\rangle,\quad
\mg^3=\tilde\mg=\mg.
$$
By Theorem~\ref{mtswe}, each transitive action $\rho$ of $\mg=\tilde\mg$ 
on a manifold of dimension $n\le 3$ determines a MT for the KdV equation, 
because the algebra $\rho(\mg^{n-1})$ is of dimension $\le n-1$ 
and cannot be transitive.

According to~\cite{cfg1},
for a transitive action $\rho\colon\mg\to D(W)$ 
the image $\rho(\mg)$ is finite-dimensional and is of the form
\begin{equation}
\label{quot}
\msl_2\otimes \Com[\la]/\bigl(f(\la)\bigl),\quad f(\la)\in\Com[\la],
\end{equation}
where $\bigl(f(\la)\bigl)$ is the ideal of $\Com[\la]$ generated by $f(\la)$.
Let
$$
f(\la)=a\prod_{s=1}^k(\la-e_s)^{k_s},\quad a,\,e_s\in\Com,\quad a\neq 0,\quad 
e_i\neq e_j\ \forall\,i\neq j.
$$
Then Lie algebra~\er{quot} is isomorphic to
\begin{equation}
\label{ds}
\bigoplus_{s=1}^k\msl_2\otimes \Com[\la]/(\la^{k_s}).
\end{equation}
The Lie groups
$\prod_{s}\mathrm{SL}_2\bigl(\Com[\la]/(\la^{k_s})\bigl)$
that appear in \cite{drin} have~\er{ds} as their Lie algebras.
Thus construction of MTs arising from this ZCR is reduced 
to local description of homogeneous spaces of $\dim\le 3$
of these Lie groups. 
This description and the corresponding MTs are presented in \cite{drin}.

A translation-invariant MT over the KdV equation 
belongs to the list of MTs in~\cite{drin} if and only if 
for the corresponding action $\rho$ of $\we$ we have $\rho(H)=0$. 

\begin{example} 
Consider the following action of $\we$ on $\Com$ 
\begin{equation*}
%\label{rc}
\mg\to 0,\quad 
	r_i\mapsto 0,\ i=-2,0,1,2,\quad 
	r_{-1}\mapsto \frac{\pd}{\pd w},
\end{equation*}
where $w$ is a coordinate in $\Com$.
%Consider the following transitive action of $\mg$ on $\Com$:
%$\msl_2\otimes\Com[\la]$ acts trivially and $H$ acts by homomorphism~\er{rc}.
The corresponding MT is 
\begin{equation*}
	u=3v_1,\quad v_t=v_3+\frac32 v_1^2
\end{equation*}	
and does not belong to the list of MTs in~\cite{drin}.
\end{example}

\subsection{MTs of a Harry-Dym type equation} 
 
Consider the equation~\cite{hdbt,konop}
\begin{equation}
\label{hd5}
u_t=u^{5/2}u_{5}
\end{equation}
in the complex-analytic category.
There are B\"acklund transformations connecting~\er{hd5} 
with the Sawada-Kotera and Kaup-Kupershmidt equations~\cite{hdbt}. 

According to~\cite{hdbt}, the Wahlquist-Estabrook algebra of~\er{hd5}
is the direct sum $\Com^2\oplus\mg$, where $\Com^2$ is a 
commutative algebra with a basis $C_1,\,C_2$ and 
$\mg$ is the ``positive part'' of the twisted affine algebra $A_2^{(2)}$. 
In other words, 
the algebra $\mg$ is isomorphic to a subalgebra of 
$\msl_3(\Com)\otimes_\Com\Com[\la]$ generated by the two elements 
\begin{equation}
\label{sl3}
X_1=
\left\lvert\!\left\lvert
\begin{array}{ccc}
0&0&0\\
\la&0&0\\
0&0&0
\end{array}\right\lvert\!\right\lvert,\quad
X_2=
\left\lvert\!\left\lvert
\begin{array}{ccc}
0&0&1\\
0&0&0\\
0&1&0
\end{array}\right\lvert\!\right\lvert
\end{equation}
satisfying the relations 
\begin{gather}
\label{hd5rel1}
(\ad X_1)^2X_2=0,\\
\label{hd5rel2}
(\ad X_2)^5X_1=0.
\end{gather}
The corresponding ZCR reads 
\begin{equation*}
A(u)=C_1u^{-\frac12}+X_1u^{-\frac32}+X_2,
\end{equation*}
the form of $B(u,u_1,u_2,u_3,u_4)$ can be found in~\cite{hdbt} 
and is not important for our purposes.

The subalgebra $\we^k,\,k\ge 1$, is generated by the elements
\begin{equation*}
C_1,\quad (\ad X_2)^iX_1,\quad i=0,\dots,k-1.
\end{equation*}
Combining this with relation~\er{hd5rel2}, 
we obtain $\we^5=\we^6=\tilde\we$. 
By Theorem~\ref{xtorder}, we obtain that any MT over~\er{hd5} 
is of order not greater than 5. 
 
Let us construct a MT over~\er{hd5}.
Consider the homomorphism $\we\to\msl_3$ that maps $C_i$ to zero 
and substitutes $\la=1$ in~\er{sl3}. Combining the standard action of $\msl_3$ 
on $\Com^3$ with this homomorphism, we obtain the following transitive action 
\begin{gather*}
\rho\colon\we\to D(\Com^3),\quad\rho(C_i)=0,\\
\rho(X_1)=w^2\frac{\pd}{\pd w^1},\quad\rho(X_2)=w^3\frac{\pd}{\pd w^2}+w^1\frac{\pd}{\pd w^3}. 
\end{gather*}
According to Theorem~\ref{mts}, to get a MT from this action we need 
to find a nonconstant function $w$ on $\Com^3$ such that $\rho(\we^2)(w)=0$. 
Since the algebra $\rho(\we^2)$ is commutative and spanned by the vector fields
\begin{equation*}
\rho(X_1)=w^2\frac{\partial}{\partial w^1},\quad
\rho([X_2,X_1])=w^3\frac{\partial}{\partial w^1}-w^2\frac{\partial}{\partial w^3},  
\end{equation*}
we can take $w=w^2$. 
By formula~\er{w} for $M(x,t,u)=A(u)$, we have 
$\tilde w^1=w^2$, $\tilde w^2=w^3$, $\tilde w^3=w^1$. 
Rewriting the vector field $\rho(A(u))$ 
in these coordinates, we obtain that 
the $x$-part of the corresponding system~\er{dsc} is
\begin{equation}
\label{hd5wx}
\frac{\partial\tilde w^1}{\partial x}=\tilde w^2,\quad 
\frac{\partial\tilde w^2}{\partial x}=\tilde w^3,\quad 
\frac{\partial\tilde w^3}{\partial x}=u^{-\frac32}\tilde w^1.
\end{equation}
Applying the substitution~\er{wv} for $w^i=\tilde w^i$, 
from~\er{hd5wx} we obtain the MT 
$$
u=\left(\frac{v}{v_3}\right)^{\frac23}.
$$  

The corresponding equation~\er{evol1} can be obtained either 
by straightforward computation or using the vector field 
$\rho(B(u,u_1,u_2,u_3,u_4))$ as described in Section~\ref{scal}. 
The answer is 
\begin{equation}
\label{hd5m}
v_t=-9vD^2\left(\left(\frac{v}{v_3}\right)^{\frac23}\right)
+\frac92v_1D\left(\left(\frac{v}{v_3}\right)^{\frac23}\right)
-\frac32v_2\left(\frac{v}{v_3}\right)^{\frac23},
\end{equation}
where $D=\sum_i v_{i+1}\pd/\pd v_i$.

\section{MTs of nonscalar evolution systems}

\label{mtsys}

\subsection{MTs from ZCRs} 
\label{zcrmtsys}

A MT~\er{mt} of nonscalar systems cannot always be written 
in some simple analog of the form~\er{dsc}, and, therefore, not all MT 
of nonscalar systems can be described by our method. 
In this section we study the MTs that can be written in a form 
analogous to~\er{dsc}. 

Consider an evolution system~\er{su} and a system 
\begin{gather}
\label{sysw}
  \frac{\pd w^j}{\pd x}=w^{i+1},\quad j=1,\dots,n-s,\\
\label{sysa}
  \frac{\pd w^{n-s+i}}{\pd x}=A^i(w^1,\dots,w^{n},x,t,u^1,\dots,u^k),\quad i=1,\dots,s,\\
\label{sysb}
  \frac{\pd w^l}{\pd t}=b^i(w^1,\dots,w^{n},x,t,u^1,\dots,u^k,u_1^1,\dots,u^k_1,\dots),\\
\notag   l=1,\dots,n.
% \end{aligned}
\end{gather}
consistent modulo~\er{su}. 
By analogy with~\er{du}, suppose that the following condition holds 
\begin{equation}
\label{sysdu}
\begin{split}
\exists\,w^1_0,\dots,w^{n}_0,\,x_0,\,&t_0,\,u^1_0,\dots,u^k_0\\
\mathrm{rank}\,\left\lvert\!\left\lvert\frac{\pd A^i}{\pd u^j}\right\lvert\!\right\lvert
(w^1_0,\dots,w^{n}_0,&x_0,t_0,u^1_0,\dots,u^k_0)=s.
\end{split}
\end{equation}
In particular, from~\er{sysdu} we have $s\le k$.

Then one gets a MT over~\er{su} as follows.
Taking into account~\er{sysdu}, by the implicit function theorem, 
there are $1\le j_1<\dots<j_s\le k$ such that $u^{j_1},\dots,u^{j_s}$ can 
locally be expressed from~\er{sysa} in terms of 
$$
x,\ t,\quad u^{n_1},\dots,u^{n_{k-s}},\quad w^1,\dots,w^{n},\quad 
\frac{\pd w^{n-s+i}}{\pd x},\quad i=1,\dots,s,
$$
where $n_1<n_2<\dots<n_{k-s}$ are such that 
$$
\{j_1,\dots,j_s,n_1,\dots,n_{k-s}\}=\{1,\dots,k\}.
$$ 
Substitute everywhere 
\begin{gather*}
w^j=v^1_{j-1},\quad \frac{\pd w^{n-s+1}}{\pd x}=v^1_{n-s},\quad j=1,\dots,n-s,\\ 
w^{n-s+j}=v^{j},\quad \frac{\pd w^{n-s+j}}{\pd x}=v^{j}_1,\quad j=2,\dots,s,\\
u^{n_j}=v^{s+j},\quad j=1,\dots,k-s.
\end{gather*}
Thus we expressed $u^i$ in terms of $x,\,t,\,v^{j}_l$, that is, we got~\er{mt}. 
Finally, system~\er{sv} is obtained from~\er{sysb}.

The total derivative operators are now 
\begin{align*}
%\label{dx}
  D_x&=\frac{\pd}{\pd x}+\sum_{\substack{i=1,\dots,k,\\ j\ge 0}} u^i_{j+1}\frac{\pd}{\pd u^i_j},\\
%\label{dt}
  D_t&=\frac{\pd}{\pd t}+\sum_{\substack{i=1,\dots,k,\\ j\ge 0}} 
D_x^j\bigl(P^i(x,t,u,u_1,\dots,u_p)\bigl)\frac{\pd}{\pd u^i_j}.
\end{align*}
Similarly to~\er{dscg} we consider systems
\begin{equation}
  \label{sysdscg}
  \begin{aligned}
  \frac{\pd w^i}{\pd x}& =a^i(w^1,\dots,w^{n},x,t,u^1,\dots,u^k),\quad i=1,\dots,n,\\
  \frac{\pd w^i}{\pd t}&=b^i(w^1,\dots,w^{n},x,t,u^1,\dots,u^k,u_1^1,\dots,u^k_1,\dots),
  \quad i=1,\dots,n.
  \end{aligned}
\end{equation}
consistent modulo \er{su} with the same equivalence relation~\er{ww}. 

Two functions
\begin{equation*}
%\label{syszcr}
M(x,t,u^1,\dots,u^k),\quad N(x,t,u^1,\dots,u^k,u_1^1,\dots,u^k_1,\dots)
\end{equation*}
with values in a Lie algebra $\mg$ constitute a ZCR for~\er{su} if equation~\er{zcr1} holds.
Each action $\rho\colon\mg\to D(W)$ and a choice of local coordinates $w^1,\dots,w^n$ 
in $W$ determine a consistent system of the form~\er{sysdscg} as follows
\begin{gather*}
    \rho(M)=\sum_{i=1}^n a^i(w^1,\dots,w^{n},x,t,u^1,\dots,u^k)\frac{\pd}{\pd w^i},\\
    \rho(N)=\sum_{i=1}^n b^i(w^1,\dots,w^{n},x,t,u^1,\dots,u^k,u_1^1,\dots,u^k_1,\dots)\frac{\pd}{\pd w^i}.
\end{gather*}

Similarly to Definition~\ref{gk},
we define subalgebras $\mg^k$ of $\mg$ by induction on $k$ as follows:
\begin{itemize}
    \item $\mg^0=0$,
    \item $\mg^1$ is the subalgebra generated by all elements 
$$
M(x,t,u^1,\dots,u^k)-M(x',t',(u^1)',\dots,(u^k)')\in\mg,
$$
\item $\mg^{k+1}$ is generated by the subspaces $\mg^k$ and 
$$
[\mg^k,M(x,t,u^1,\dots,u^k)].
$$
\end{itemize}
Set also $\tilde\mg=\cup_{k\ge 0}\mg^k$.

\begin{theorem}
\label{sysmts} 
Suppose that system~\er{sysdscg} arises from an action $\rho\colon\mg\to D(W)$
and there are $x_0,\,t_0,\,u^1_0,\dots,u^k_0$, 
$z\in W$ and a neighborhood $W_0$ of $z$ such that 
\begin{gather*}
\label{syszmn}
\dim\ev_{\rho,z}(\tilde\mg)=n,\\
\label{ans}
\forall\,a\in W_0\quad \dim\ev_{\rho,a}(\mg^{n-s})<n,
\end{gather*}
where $s$ is the dimension of the subspace
\begin{equation}
\label{mtz}
\left.\left\langle\frac{\pd\rho(M)}{\pd u^i}(z,x_0,\,t_0,\,u^1_0,\dots,u^k_0)\,\right\lvert\,
i=1,\dots,k\right\rangle\subset T_zW.
\end{equation}
Then on a neighborhood of $z$ there is a function $w$ such that 
\begin{equation*}
%\label{sysdzw}
d_zw\neq 0,\quad \rho(\mg^{n-s})(w)=0,
\end{equation*}
and it is unique up to a change $w\mapsto g(w)$. The functions
\begin{equation*}
%\label{sysww}
\tilde w^i=\rho\bigl(M(x,t,u^1,\dots,u^k)\bigl)^{i-1}(w),\quad i=1,\dots,n-s+1,
\end{equation*}
do not depend on~$x,\,t,\,u^i$. 
One can find functions $\tilde w^{n-s+2},\dots,\tilde w^{n}$ 
such that $\tilde w^1,\dots,\tilde w^n$ form a system of coordinates 
on a neighborhood of $z$. 
The initial system~\er{dscg} takes 
the form~\er{sysw},~\er{sysa},~\er{sysb},~\er{sysdu} in these coordinates.
\end{theorem}
\begin{proof}
Similarly to Theorem~\ref{mts}, this follows from Lemma~\ref{ooo} 
for $V=\rho\bigl(M(x,t,u^1,\dots,u^k)\bigl)$. 
Condition $m_1=\dim\ev_{\rho,z}(\mg^1)\ge s$ 
holds because the space $\ev_{\rho,z}(\mg^1)$ includes subspace~\er{mtz}.
\end{proof}

\subsection{MTs of the Kersten-Krasilshchik system}

The following system was introduced in~\cite{kdv-mkdv} and proved to be integrable
\begin{equation}
\label{kdvmkdv}
\begin{aligned}
u^1_t&=-u^1_3+6u^1u^1_1-3u^2u^2_3-3u^2_1u^2_2+3u^1_1(u^2)^2+6u^1u^2u^2_1,\\
u^2_t&=-u^2_3+3(u^2)^2u^2_1+3u^1u^2_1+3u^1_1u^2.
\end{aligned}
\end{equation}

Denote by $E_{ij},\,i,j=1,2,3$, the $(3\times 3)$-matrix with $(i,j)$-entry equal to $1$  
and other entries equal to $0$. 
Let $\mg$ be the $5$-dimensional Lie subalgebra of $\msl_3(\Com)$ spanned by the elements 
$$
e=E_{12},\quad n_1=E_{13},\quad f=E_{21},\quad n_2=E_{23},\quad h=E_{11}-E_{22}.
$$
An $\msl_3$-valued ZCR dependent on a parameter $\lambda$ 
was constructed for~\er{kdvmkdv} in~\cite{kdv-mkdv-sak}. 
For $\la=0$ one obtains the following ZCR
\begin{equation*}
M=((u^2)^2-u^1)e-f-u^2n_1
\end{equation*}
and 
\begin{multline*}
N=(u^1_2+u^2u^2_2+(u^2_1)^2-2(u^1)^2+(u^2)^4+(u^2)^2u^1)e\\
-(u^1_1+u^2u^2_1)h-((u^2)^2+2u^1)f+(u^2_2-(u^2)^3-2u^1u^2)n_1+u^2_1n_2.
\end{multline*}
Let us construct MTs for~\er{kdvmkdv} from this ZCR using Theorem~\ref{sysmts}.
We have 
\begin{gather}
\label{g1}
\mg^1=\langle e,\,n_1\rangle,\\ 
\notag
\mg^2=\langle h,\,e,\,n_1,\,n_2\rangle,
\quad\mg^3=\tilde\mg=\mg.
\end{gather}

Every subalgebra $\mh\subset\mg$ of codimension $3$ such that 
\begin{equation}
\label{gh}
\mg^1\cap\mh=0
\end{equation} 
determines a MT as follows. Consider Lie groups $H\subset G$ corresponding 
to the Lie algebras $\mh\subset\mg$ and set $W=G/H$, $z=H\in G/H$.   
We have the natural transitive action of $\rho\colon\mg\to D(W)$. 
From~\er{g1} and~\er{gh} we obtain that the dimension of space~\er{mtz} 
is equal to $2$. Since $\dim\mg^1=2$ and $\dim W=3$, the algebra $\rho(\mg^1)$ 
is not transitive on $W$. Therefore, all conditions of Theorem~\ref{sysmts} 
are satisfied, and $\rho$ determines a MT with $n=3$, $s=2$. 

For example, set $\mh=\langle n_2,\,f-n_1\rangle$. 
The corresponding action in local coordinates reads
\begin{gather*}
e\mapsto\frac{\pd}{\pd w^2},\quad n_1\mapsto \frac{\pd}{\pd w^3},\quad n_2\mapsto w^2\frac{\pd}{\pd w^3},\\
f\mapsto w^2\frac{\pd}{\pd w^1}-(w^2)^2\frac{\pd}{\pd w^2}+(\mathrm{e}^{-3w^1}-w^2w^3)\frac{\pd}{\pd w^3},\\ 
h\mapsto  \frac{\pd}{\pd w^1}-2w^2\frac{\pd}{\pd w^2}-w^3\frac{\pd}{\pd w^3}.
\end{gather*}
Indeed, the isotropy subalgebra of the point $z=(0,0,0)$ is equal to $\mh$. 
The MT reads 
\begin{gather*}
u^1=v^1_2+(v^1_1)^2+(v^2_1+v^1_1v^2-\mathrm{e}^{-3v^1})^2,\\
u^2=v^2_1+v^1_1v^2-\mathrm{e}^{-3v^1}.
\end{gather*}
The corresponding system~\er{sv} is cumbersome and can be obtained from 
$\rho(N)$ as described in Subsection~\ref{zcrmtsys}.

\subsection*{Acknowledgements} 
 
The author thanks V.~V.~Sokolov for useful discussions.

\end{document}